\documentclass[aps, reprint, 10pt, superscriptaddress, twocolumn, prb]{revtex4-1}
\usepackage{hyperref}
\usepackage{float}
\usepackage{longtable}
\usepackage{mathtools}
\usepackage[utf8]{inputenc}
\usepackage[T1]{fontenc}
\usepackage{lmodern}
\usepackage{url}
\usepackage[table]{xcolor}
\usepackage{multirow}
\usepackage{siunitx}
\usepackage{booktabs}
\begin{document}

\title{Anomalous Non-Hydrogenic Exciton Series in 2D Materials on High-$\kappa$ Dielectric Substrates}

\author{Anders C. Riis-Jensen}
\affiliation{Center for Atomic-scale Materials Design, Department of
Physics, Technical University of Denmark, DK - 2800 Kongens Lyngby,
Denmark}
\author{Morten N. Gjerding}
\affiliation{Center for Atomic-scale Materials Design, Department of
Physics, Technical University of Denmark, DK - 2800 Kongens Lyngby,
Denmark}
\author{Saverio Russo}
\affiliation{Centre for Graphene Science, College of Engineering, Mathematics and Physical Sciences, University of Exeter, Exeter EX4 4QL, UK}
\author{Kristian S. Thygesen \email{Electronic address:
thygesen@fysik.dtu.dk}}
\affiliation{Center for Atomic-scale Materials Design, Department of
Physics, Technical University of Denmark, DK - 2800 Kongens Lyngby,
Denmark}
\email{thygesen@fysik.dtu.dk}
\date{\today}

\begin{abstract}
Engineering of the dielectric environment represents a powerful strategy to control the electronic and optical properties of two-dimensional (2D) materials without compromising their structural integrity. Here we show that the recent development of high-$\kappa$ 2D materials present new opportunities for dielectric engineering. By solving a 2D Mott-Wannier exciton model for WSe$_2$ on different substrates using a screened electron-hole interaction obtained from first principles, we demonstrate that the exciton Rydberg series changes qualitatively when the dielectric screening within the 2D semiconductor becomes dominated by the substrate. In this regime, the distance dependence of the screening is reversed and the effective screening increases with exciton radius, which is opposite to the conventional 2D screening regime. Consequently, higher excitonic states become underbound rather than overbound as compared to the Hydrogenic Rydberg series. Finally, we derive a general analytical expression for the exciton binding energy of the entire 2D Rydberg series  . 
\end{abstract}

\maketitle
Intense research during the past decade has established the unique optical properties of two-dimensional (2D) semiconductors such as single layers of transition metal dichalcogenides (MX$_2$ and MXY with M=Mo,W and X=S,Se,Te)\cite{mak2010atomically,cheiwchanchamnangij2012quasiparticle,zhao2013evolution,molina2016temperature,riis2019classifying}. Most fundamentally, these ultrathin materials host strongly bound excitons with binding energies reaching up to 25\% of the band gap\cite{ramasubramaniam2012large,qiu2013optical,zhang2014absorption} making them candidates for excitonic devices that can be operated at room temperature. Another unique feature is the extreme degree to which these excitons can be manipulated and controlled externally, e.g. via the dielectric environment\cite{ruiz2019interlayer,jin2019observation,trolle2017model, borghardt2017engineering,park2018direct,raja2017coulomb,steinhoff2018frequency,lin2014dielectric}. Higher-lying excitonic states also exhibit unusual properties. In particular, the Rydberg series does not follow the $1/n^2$ Hydrogenic series known from 3D materials, but show a distinct non-Hydrogenic behavior as a direct consequence of the non-local nature of the 2D dielectric function\cite{chernikov2014exciton,liu2019magnetophotoluminescence}. For a translation invariant system the $q$-dependent dielectric function obeys  
\begin{equation}
    \epsilon(\mathbf q)= \int d(\mathbf{r}-\mathbf{r}')\epsilon(\mathbf{r}-\mathbf{r}')e^{i\mathbf{q}\cdot(\mathbf{r}-\mathbf{r}')}
\end{equation}
From this it follows that non-locality in real space translates into $q$-dependence in reciprocal space. In bulk semiconductors it is typically a good approximation to set $\epsilon(\mathbf r-\mathbf r')=\epsilon\delta(\mathbf r-\mathbf r')$, which yields a $q$-independent dielectric function. In contrast, for a freestanding 2D semiconductor $\epsilon(q)=1+\alpha q$ (for small $q$), from which it follows that screening in 2D is notoriously non-local.  
Furthermore, it has been shown that the dielectric function of a 2D material is sensitive to its dielectric environment, e.g. a substrate. However, with a few notable exceptions\cite{ugeda2014giant,qiu2019giant}, all experiments on excitons in 2D semiconductors reported to date employed substrates with weak dielectric screening, e.g. hBN, SiO$_2$ or sapphire\cite{komsa2012effects,hsu2019dielectric,yan2018superior,qiu2017environmental}. As a consequence, the developed theory of excitons in atomically thin materials have also exclusively focused on this scenario.  

In this work, we explore what happens to the excitonic states of a 2D semiconductor, here exemplified by WSe$_2$, when the dielectric screening inside the 2D material becomes dominated by the environment. For WSe$_2$ on low-screening substrates we obtain the well known non-hydrogenic exciton series\cite{chernikov2014exciton,olsen2016simple} where higher exciton states are screened less than the $n=1$ ground state. Here we show that this trend is reversed when the dielectric screening from the substrate exceeds the intrinsic screening in the 2D layer. At the end we generalize the 2D hydrogen-like model previously developed\cite{olsen2016simple} for the exciton effective dielectric constant in a 2D semiconductor to include the effect of neighbouring substrate screening and obtain excellent agreement with numerical solutions.

\begin{figure}[h!]
\centering
\includegraphics[scale=0.20]{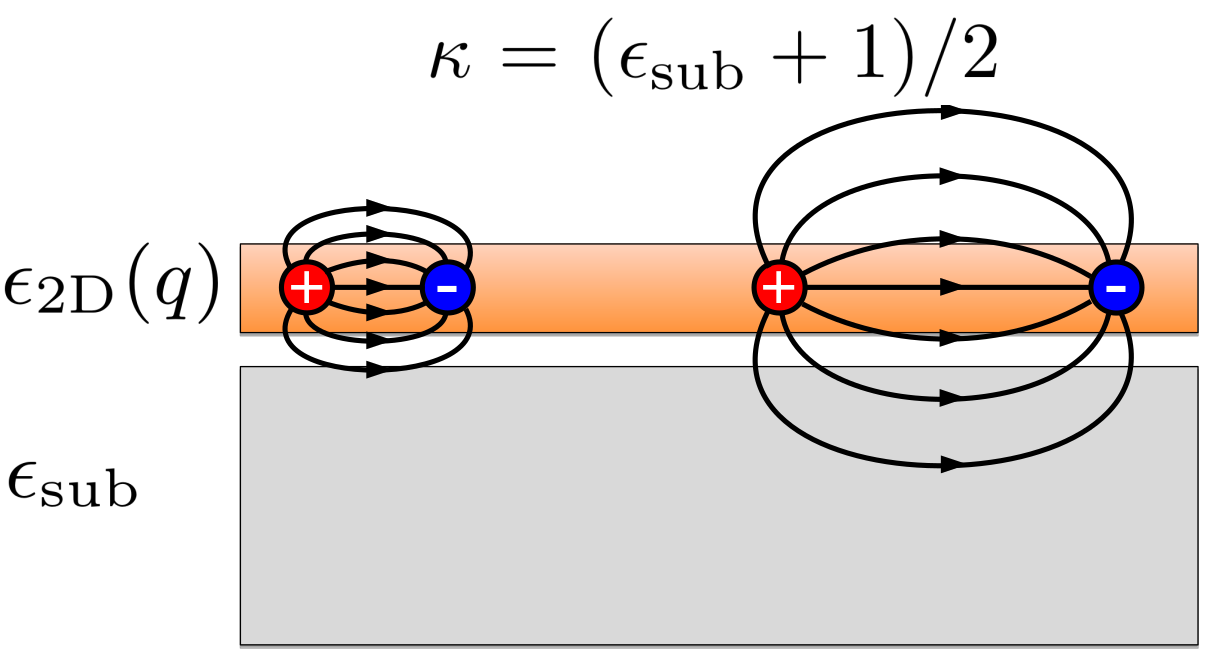}
\caption{A 2D layer with non-local, i.e. $q$-dependent, dielectric function located on a dielectric media with constant permittivity. 
The extension of the electric fields lines from two 2D excitons with different spatial radius is depicted. The larger the exciton radius, the more the screening is determined by the environment outside the 2D material.}\label{fig:setup}
\end{figure}

The excitonic states of the WSe$_2$ monolayer are obtained by solving the Mott-Wannier equation for the exciton envelope wave function
\begin{equation}\label{eq:mottwannier}
    \left[ - \frac{\hbar^2}{2\mu}\left(\frac{\partial^2}{\partial x^2}+\frac{\partial^2}{\partial y^2}\right) + W(\textbf{r}) \right] F(\textbf{r}) = E_n F(\textbf{r})
\end{equation}
where $\mu$ is the exciton effective mass, $W(\textbf{r})$ is the screened electron-hole interaction, $E_n$ the binding energy of state $n$, and $F(\textbf{r})$ is the probability amplitude for finding the electron and hole at separation $\mathbf{r}$. We use an exciton mass for WSe$_2$ of $\mu=0.187$ adopted from the Computational 2D Materials Database\cite{haastrup2018computational} 
To calculate the screened interaction between the electron and hole in the 2D layer we use the Quantum Electrostatic Heterostructure (QEH) model\cite{andersen2015dielectric}, which has previously been shown to yield accurate energies for excitons in van der Waals heterostructures\cite{riis2018efficient, latini2017interlayer}. The QEH model takes the \emph{ab-initio} response functions of the 2D layer as input and thus accounts fully for the non-local screening in the monolayer. Screening from the environment is taken into account by the method of image charges. We define the effective permittivity of the environment as $\kappa=(\epsilon_\mathrm{sub}+1)/2$, where $\epsilon_\mathrm{sub}$ and 1 are the permittivities of the half spaces above and below the 2D material, see Fig. \ref{fig:setup}.  We here restrict our calculations to merely consider a 2D monolayer on a dielectric substrate. We do this since by the virtue of the defintion of $\kappa$, the results obtained for a 2D monolayer on a dielectric substrate with dielectric constant $\epsilon_a$, will be quantitatively very close to the same 2D monolayer encapsulated between two substrates, both with dielectric constants $\epsilon_a / 2$, and therefore this method is straightforward to generalize to the case of an encapsulated 2D monolayer. All future calculations and references for substrates therefore addresses the setup in Fig. \ref{fig:setup} with a a substrate on one side and vacuum on the other side. 

\begin{figure}[h!]
\centering
\includegraphics[scale=0.40]{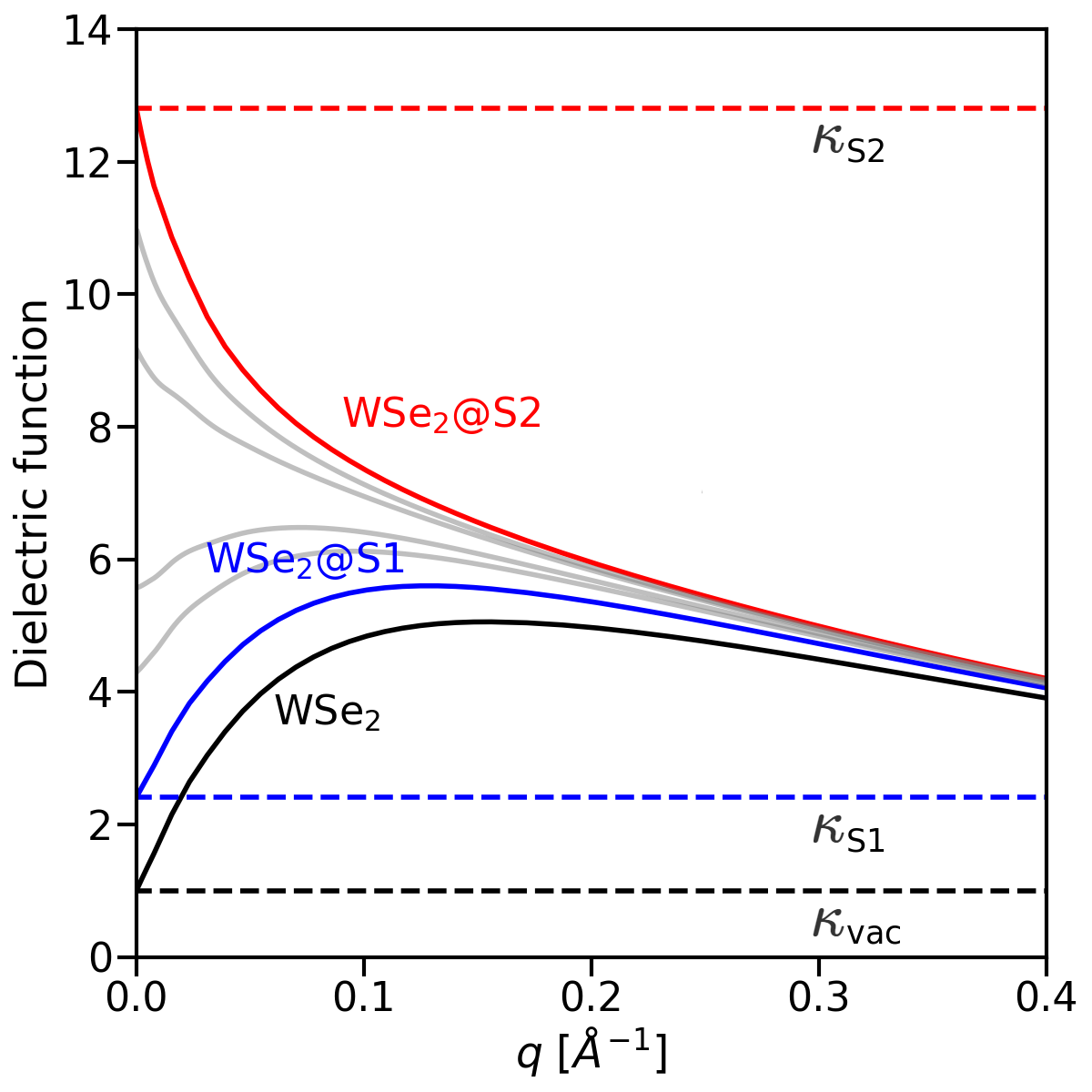}
\caption{The $\textbf{q}$-dependent dielectric function of monolayer WSe$_2$ in isolation (black), in a weakly screening environment (blue), and in a strongly screening environment (red). In the limit $q\to 0$, the dielectric function becomes equal to the constant permittivity of the environment, $\kappa$. 
When the screening from the environment exceeds the intrinsic screening in WSe$_2$, the slope of $\epsilon(q)$ at $q=0$ changes from positive to negative. The light gray lines show intermediate systems illustrating the transition from one screening regime to the other.}\label{fig:eps_q}
\end{figure}

Before turning to the results of the exciton calculations, we discuss how the dielectric function of the WSe$_2$ monolayer is influenced by the dielectric environment. We define the dielectric function of the 2D monolayer by
\begin{equation}
    \epsilon(q) \equiv \frac{V(q)}{W(q)}
\end{equation}
where $V(q)=1/q$ is the 2D Fourier transform of $1/r$ and $W(q)$ is the screened Coulomb interaction between two point charges in the 2D layer as obtained from the QEH model. In Fig. \ref{fig:eps_q} we show the dielectric function of freestanding WSe$_2$ (black), WSe$_2$ on a weakly screening substrate (blue, denoted S1), and a strongly screening substrate (red, denoted S2). The horizontal dashed lines ($\kappa_\mathrm{2D}$, $\kappa_\mathrm{S1}$, and $\kappa_\mathrm{S2}$) mark the $q = 0$ limits of the dielectric functions, which equal the effective permittivity of the environment (see below). We note that, $\kappa_\mathrm{S2} = 13$, could easily be realized upon encapsulation of WSe$_2$ in the high-layered materials HfOx\cite{de2018strain,peimyoo2019laser} , while the $\kappa_\mathrm{S1} = 2.5$ corresponds to WSe$_2$ on a hBN substrate. 

We can rationalize the small-$q$ behaviour of the dielectric functions in Fig. \ref{fig:eps_q}, by 
combining two basic facts about screening in 2D. Firstly, in the small $q$-limit, the density response function of a semiconductor takes the form, $\chi^0(q)=-\alpha q^2$ (independent of dimensionality). Secondly, the electrostatic potential from a 2D charge distribution of the form $\rho(\mathbf r,z)= e^{i \textbf{q} \cdot \textbf{r}}\delta(z)$, equals
\begin{equation}\label{eq:expo}
    V(\textbf{r},z) = \frac{1}{q} e^{i \textbf{q} \cdot \textbf{r}} e^{-q |z|}. 
\end{equation}\label{eq:V}
Focusing first on the case of an isolated 2D layer, i.e. ignoring the $z$-dependence, the dielectric function becomes $\epsilon_\mathrm{2D}(q)=1-V(q)\chi^0(q)=1+\alpha q$. In particular, there is no intrinsic screening from the 2D layer in the $q\to 0$ limit. Next, consider the effect of a substrate. It should be clear from Eq. (\ref{eq:expo}), that in the $q=0$ limit all screening is due to the environment and thus $\epsilon(0)=\kappa$ (potentials do not decay away from the layer and there is no intrinsic screening from the layer itself). Moreover, the contribution to the dielectric function from the environment will be exponentially suppressed for larger $q$. These considerations are evidently in agreement with the results in Fig. \ref{fig:eps_q}. Interestingly, when $\kappa$ becomes comparable to the maximum permittivity of the 2D layer, $\epsilon(q)$ changes qualitatively; in particular, the slope at $q=0$ changes from positive to negative.

\begin{figure}[h!]
\centering
\includegraphics[scale=0.27]{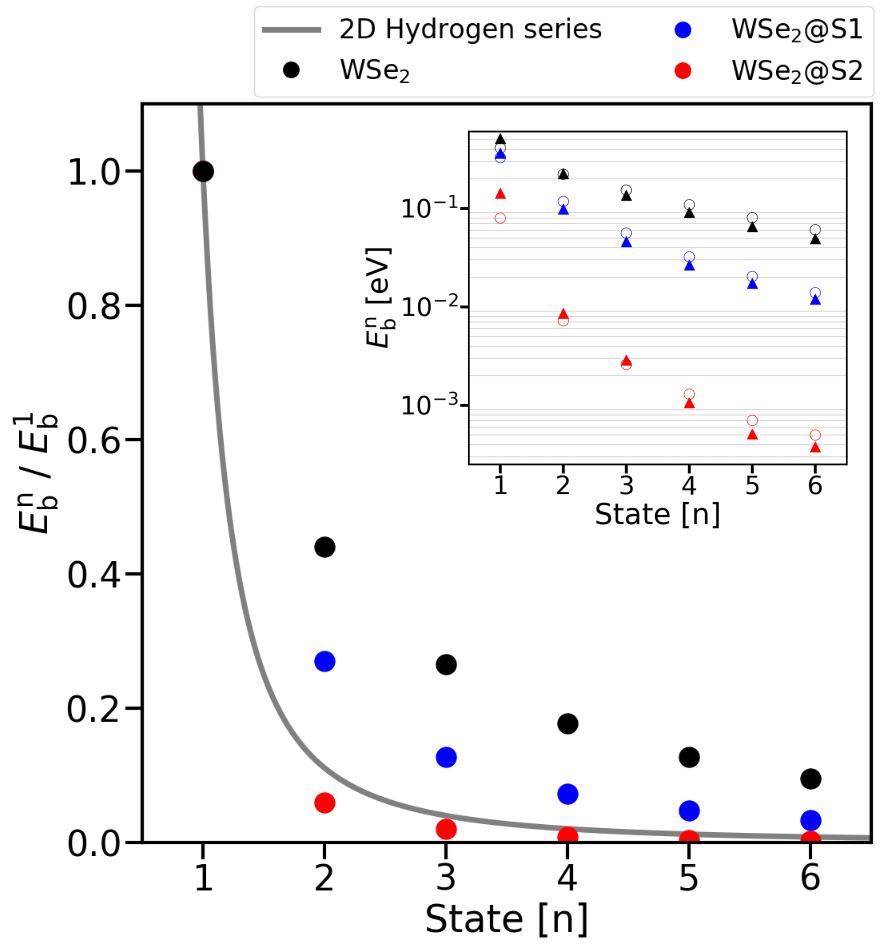}
\caption{Exciton Rydberg series of WSe$_2$ plotted relative to the $n=1$ state. The different colors represent the exciton binding energies obtained from the Mott-Wannier model with environmental screening corresponding to the three dielectric functions in Fig. \ref{fig:eps_q}. The universal Hydrogenic series is shown by the grey curve. The inset shows the actual exciton binding energies for the Mott-Wannier model (triangles) and from Eq. \ref{eq:Eb_hyd}, with the analytical expression for the effective screening, Eq. \ref{eq:effeps_analytical} (empty circles)}.\label{fig:eb_series}
\end{figure}

At this point we return to the Mott-Wannier exciton model Eq. (\ref{eq:mottwannier}). It is well known that the exciton Rydberg series in a 2D semiconductor is distinctly different from the usual Hydrogenic series observed in 3D bulk materials\cite{chernikov2014exciton}. This can be understood as a direct consequence of $\epsilon(q)$ being an increasing function of $q$ in the relevant wavevector range from 0 to around $1/a$ where $a$ is a characteristic exciton size ($a>10$ \AA) for the TMDs. This form of $\epsilon(q)$ results in excitons with higher $n$ being less screened due to their more delocalized nature\cite{olsen2016simple} (more delocalized in real space corresponds to more localized in reciprocal space). This trend is reflected by the black symbols in Fig. \ref{fig:eb_series}, which shows the binding energies obtained from Eq. (\ref{eq:mottwannier}) for the lowest ($l=0$) exciton states of freestanding WSe$_2$ normalized to the $n=1$ state. As a reference, the grey curve shows the result for a 2D Hydrogen model
\begin{equation}\label{eq:Eb_hyd}
E_\mathrm{n}^\mathrm{Hydrogen} = - \frac{\mu}{2 \left(n-\frac{1}{2} \right)^2 \kappa^2 }.
\end{equation}
Note that the constant permittivity, $\kappa$, and exciton mass, $\mu$, do not enter the scaled binding energy, $E_\mathrm{n}^\mathrm{Hydrogen}/E_\mathrm{1}^\mathrm{Hydrogen}$, which can thus be taken as a universal curve corresponding to the situation of local screening. 

When WSe$_2$ is placed in a weakly screening environment, e.g. on an hBN substrate corresponding to the blue curve in Fig. \ref{fig:eps_q}, the scaled exciton binding energies move closer to the Hydrogenic series but still display the same trend of higher excitonic states being underscreened relative to the $n=1$ ground state. In contrast, when WSe$_2$ is placed in a strongly screening environment, e.g. encapsulated in HfO$_x$ corresponding to the red curve in Fig. \ref{fig:eps_q}, the exciton binding energies move below the Hydrogenic series. This marks a new screening regime in which the excitons become more efficiently screened the larger $n$.

\begin{figure}[h!]
\centering
\includegraphics[scale=0.20]{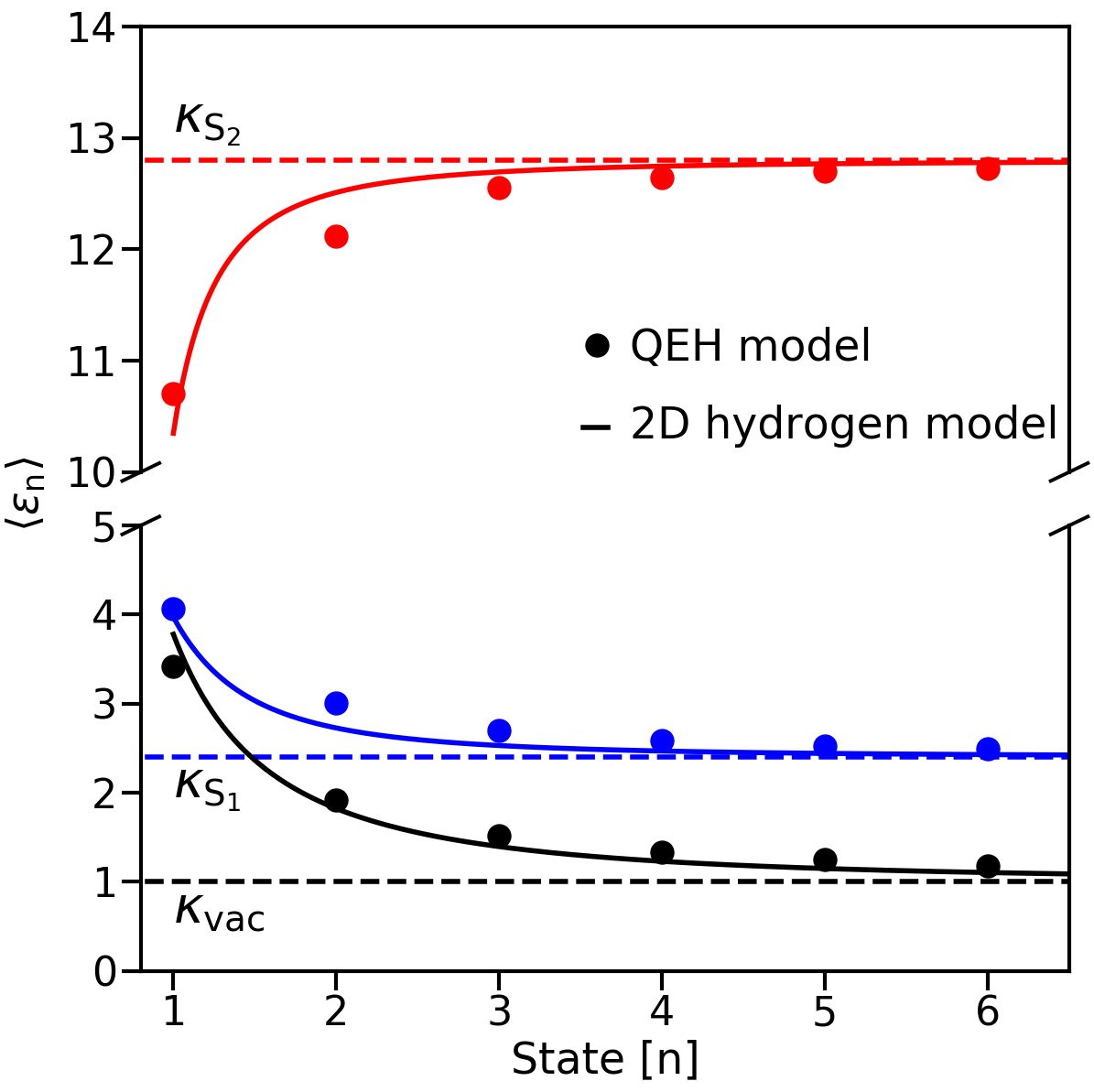}
\caption{Effective state-dependent dielectric constant for the excitons in the Rydberg series of WSe$_2$ (circular dots), see definition in Eq. \ref{eq:eff_eps}, and from the analytical solution (full lines), see definition in Eq. \ref{eq:effeps_analytical}. The different colors correspond to WSe$_2$ in different screening environments and the colour coding follows the previous figures. For $n\to \infty$, the effective dielectric constants converge towards the permittivity of the environment, $\kappa$.}\label{fig:effeps_n}
\end{figure}

To understand the transition to the new exciton screening regime, we calculate the exciton wave function for the three cases studied in Figs. \ref{fig:eps_q} and \ref{fig:eb_series} and from this extract the exciton radius for the first six states in the Rydberg series. We find an increasing exciton radius for higher states in the Rydberg series as well known from the hydrogen atom model. For the $n=1$ state of the freestanding monolayer, we obtain an exciton radius around 16 $\AA$ in good agreement with previous results\cite{latini2015excitons}. We can define an effective dielectric constant for the $n$th exciton state by averaging the $q$-dependent dielectric function over a disc with radius $1/a_n$
\begin{equation}\label{eq:eff_eps}
    \left< \epsilon_\mathrm{n} \right> = 2 a_\mathrm{n}^2 \int_{0}^{1/a_\mathrm{n}} dq\ q \epsilon(q).
\end{equation}
The result is shown in Fig. \ref{fig:effeps_n} represented by the circular dots. As the exciton radius increases monotonically with $n$ the size of the averaging disc shrinks and the effective dielectric constant $\left< \epsilon_\mathrm{n} \right>$ increases or decreases with $n$ depending on the sign of $d\epsilon/dq$ at $q=0$.

It is instructive to supplement the reciprocal-space analysis by a real space picture. As shown in Fig. \ref{fig:setup}, an increasingly larger fraction of the field lines between the electron and hole will pass through the environment as the exciton radius increases. Therefore, as $n$ increases the effective screening will change from being dominated by the 2D layer to being dominated by the environment. Consequently, whether screening of the exciton will increase or decrease with $n$ is determined by the permittivity of the environment relative to the intrinsic permittivity of the 2D layer. 

We now derive an analytical expression for the effective dielectric constant determining the screening of the exciton. We do this by generalizing the previously developed screened Hydrogen model developed in Ref. \cite{olsen2016simple} for freestanding 2D layers, to include the dielectric screening from a substrate. First we note that the exciton wave functions in reciprocal space, i.e. $F(q)$, are typically confined to small $q$-values where the intrinsic dielectric function of the 2D layer is linear. We therefore take $\epsilon(q) \approx 1 + 2 \pi \alpha q + \kappa e^{-2dq}$, where $\alpha$ is the 2D static polarizability and $d$ is the distance between the center of the 2D layer and the surface of the substrate (the factor 2 accounts for the distance to the image charge). The linear term describes the intrinsic screening from the 2D semiconductor while the last term is the substrate screening which decays exponentially away from the substrate as discussed previously. Since the exciton radius ($a_\mathrm{n}$) is itself a function of the effective exciton dielectric constant\cite{olsen2016simple},  Eq. (\ref{eq:eff_eps}) has to be solved self-consistently with the proposed form of $\epsilon(q)$. The integration can be readily carried out analytically, however to obtain a closed analytical form we Taylor expand the exponential term around $d/a_\mathrm{n}$ after integration, which is in general a good approximation for the the integration limits in eq. \ref{eq:eff_eps}. Given the large spatial extension of the exciton wave function for the higher Rydberg states the accuracy of this approximation increases with $n$. To relate $\left< \epsilon_\mathrm{n} \right>$ and $a_\mathrm{n}$ we use the relation from the ideal 2D hydrogen model with angular momentum quantum number $l=0$ (see for instance Ref.\cite{olsen2016simple}). Combining this we arrive at an expression for the effective exciton dielectric constant:
\begin{equation}\label{eq:effeps_analytical}
    \left< \epsilon_\mathrm{n} \right> = \frac{1}{2}(1 + \kappa) \left( 1 + \sqrt{1 + \frac{8 \mu \left( \frac{4}{3} \pi \alpha - \kappa d \right)}{(3n(n-1)+1)(1+\kappa)^2}} \right).
\end{equation}
The result plotted in Fig. \ref{fig:effeps_n} (full lines) shows good agreement with the QEH model for both low- and high-$\kappa$ substrates and demonstrating that the analytical expression captures the combined effect of intrinsic 2D and the substrate screening. We stress that the analytical Eq. (\ref{eq:effeps_analytical}) contains no free parameters, but is completely defined by $\alpha$, $\kappa$, and $d$. The exciton binding energy can be obtained from the 2D Hydrogen model, Eq. \ref{eq:Eb_hyd}, by replacing the $\kappa$ by the effective exciton dielectric constant. Except for underestimating  the exciton binding energy of the 1s state for the 2D and 2D@S2 systems of about 100 meV and 50 meV respectively, we find excellent agreement for all other Rydberg states, generally within about 15 $\%$ compared to the QEH results, as shown in the inset of Fig. \ref{fig:eb_series}. The underestimated value of the exciton binding energy for the 1s state was also found in the original work on freestanding 2D monolayers\cite{olsen2016simple} and can be ascribed to the larger extent of $F(q)$ for the 1s state which reduces the accuracy of the linear approximation to the intrinsic 2D screening, $\epsilon_{2D}(q)\approx 1 + 2 \pi \alpha q$. In fact this approximation overestimates the intrinsic screening for larger $q$ leading to an underestimation of the binding energy for the spatially localized 1s exciton.

Finally, we comment on the absolute size of the exciton binding energies. While the first few states of the Rydberg series of WSe$_2$ on the weakly screening substrate ($\kappa=2.5$) are at least 100 meV, the $1s$ state in the strongly screening environment ($\kappa=13$) has a binding energy around 140 meV which is reduced by more than a factor 10 for the 2s state, making it unstable at room temperature. By inserting 3 (or more) layers of hBN ($\epsilon_{\mathrm{hBN}}=4$) between WSe$_2$ and the substrate, the binding energy of the 2S state increases to 40 meV, making is stable at room temperature. While the exciton binding energies are significantly increased, the system remains in the anomalous screening regime.

In conclusion, we have identified a new screening regime for 2D semiconductors, which arises when the 2D material is placed in a dielectric environment with a permittivity exceeding that of the 2D layer itself. This anomalous screening regime is characterized by a non-local 2D dielectric function, $\epsilon(q)$, which decreases monotonically with $q$. As a consequence, whereas the usual non-Hydrogenic 2D exciton Rydberg series is characterized by states of higher $n$ being less screened and therefore stronger bound as compared to the Hydrogen series, the opposite trend is observed in the anomalous 2D screening regime. The new screening regime presents new opportunities for advancing our understanding and ability to control exciton physics in 2D semiconductors.

\twocolumngrid
\begin{acknowledgments}
The Center for Nanostructured Graphene (CNG) is sponsored by the Danish Research Foundation, Project DNRF103. The project has received funding from the European Research Council (ERC) under the European Union’s Horizon 2020 research and innovation programme (Grant No. 773122, LIMA). SR acknowledges financial support by RC-UK EPSRC (EP/K010050/1 and EP/L015331/1) and the Leverhulme Trust (Research grant: Quantum revolution).

\end{acknowledgments}

\newpage
\bibliography{paper}

\end{document}